\documentclass[preprint,showpacs,preprintnumbers,amsmath,amssymb]{revtex4}

\usepackage{graphicx}
\usepackage{dcolumn}
\usepackage{bm}
\usepackage{hyperref}

\newcommand{\qed}{\nobreak \ifvmode \relax \else
      \ifdim\lastskip<1.5em \hskip-\lastskip
      \hskip1.5em plus0em minus0.5em \fi \nobreak
      \vrule height0.75em width0.5em depth0.25em\fi}

\begin{document}

\preprint{}

\title{Comparative Bi-stochastizations and Associated 
Clusterings/Regionalizations of the 1995-2000 U. S. Intercounty Migration Network}
\author{Paul B. Slater}%
\email{slater@kitp.ucsb.edu}
\affiliation{%
University of California, Santa Barbara, CA 93106-4030\\
}%
\date{\today}

\begin{abstract}
Wang, Li and K{\"o}nig have recently compared the cluster-theoretic properties of bi-stochasticized symmetric data similarity (e. g. kernel) matrices, produced by minimizing two different forms of Bregman divergences. We extend their investigation to non-symmetric matrices, specifically studying the 1995-2000 U. S. $3,107 \times 3,107$ intercounty migration matrix. A particular bi-stochastized form of it had been obtained  (arXiv:1207.0437), using the well-established Sinkhorn-Knopp (SK) (biproportional) algorithm--which minimizes the Kullback-Leibler form of the divergence. This matrix has but a single
entry equal to (the maximal possible value of) 1. Highly contrastingly, the bi-stochastic matrix obtained here, implementing the 
Wang-Li-K{\"o}nig-algorithm for the minimum of the alternative, squared-norm form of the divergence, has 2,707 such unit entries. The  corresponding 
3,107-vertex, 2,707-link directed graph has 2,352 strong components. These consist of 1,659 single/isolated counties, 654 doublets (thirty-one interstate in nature), 22 triplets (one being interstate), 13 quartets (one being interstate), three quintets and one septet. Not manifest in these graph-theoretic results, however, are the five-county states of Hawaii and Rhode Island and the eight-county state of Connecticut. These--among other regional configurations--appealingly emerged as well-defined entities in the SK-based strong-component hierarchical clustering.
\end{abstract}


\maketitle
\section{Introduction}
We have widely applied a certain two-stage (double-standardization \cite{mosteller}, followed by strong-component hierarchical clustering \cite{tarjan2}) algorithm to (in general, non-symmetric) matrices/networks of interareal migration flows for numerous nations--as well as other forms 
of "transaction flows", of frequent occurrence in the social sciences \cite{PBSPNAS,dubes,brams}. The first stage--intended to control for multiscale effects \cite{mosteller,SBV,SBV2}--involves transforming the $n \times n$ matrix of flows to bi-stochastic form (that is, constraining all row and column sums to equal 1). To accomplish this, we have  utilized the well-known ("Sinkhorn-Knopp" [SK], as well as many other designated-names) "biproportional scaling" method 
\cite{knight}, applied in diverse areas of application. Rows and columns of the input/raw matrix are alternately scaled to sum to 1, until convergence--under broad conditions \cite{knight,hartfiel}--is obtained. This "iterative proportional fitting procedure" has a maximum-entropy-related (Bregman-divergence [Kullback-Leibler] minimization) interpretation \cite{eriksson}.

Recently,  Wang, Li and K{\"o}nig \cite{WLK} (also \cite{WLK2}, with Wan as additional co-author) have studied the use of an alternative (Euclidean-distance-based/squared-norm Bregman divergence \cite{dhillon}) procedure for bi-stochastization. (An optimization problem using the strictly convex function $\phi(x)= \frac{x^2}{2}$ is solved, while the SK-procedure is based upon the strictly convex function 
$\phi(x) =x \log{x}-x$.) They found it generally superior in terms of cluster-analytic properties to the more standard SK-based methodology (as well as to the Normalized Cut [Ncut] algorithm \cite{shimalik}).

\begin{figure}
\includegraphics{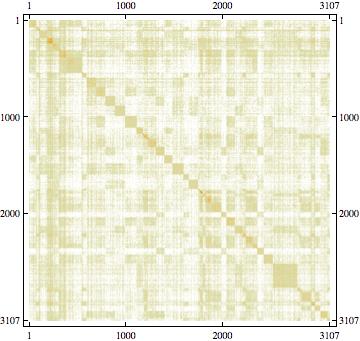}
\caption{\label{fig:MigrationMatrix}The 1995-2000 U. S. intercounty migration matrix--prior to 
bi-stochastization. The counties are ordered alphabetically by state, and by counties within states. This ordering accounts for the appearance here--and also in the next two figures--of relatively highlighted blocks, both diagonal and off-diagonal. The two states with the most counties, Texas (254) and Georgia (169), are evident.}
\end{figure}

Here, we apply this recently-developed, alternative  divergence-minimization bi-stochastization procedure to the (non-symmetric) $3,107 \times 3,107$ matrix of 1995-2000 U. S. intercounty migration (Fig.~\ref{fig:MigrationMatrix}). Then, we compare the concomitant cluster-analytic results obtained to those--based on the  SK method--previously obtained, extensively reported in 
\cite{DendrogramRegionalization} and papers cited there. Very striking differences, indeed, emerge in the two sets of results generated.  
In none of the clusterings discussed are contiguity constraints {\it a priori} imposed. Nevertheless, the results, to a very large extent, manifest contiguous effects. \footnote{In an interesting recent study, the authors sought to determine "space-independent communities in spatial networks"--first having removed distance (cf. contiguity-related) effects \cite{EEBL}.}

Michael Trott of Wolfram Research  provided us with advanced Mathematica code Supplemental Material sec.~A 
for the squared-norm-bi-stochastization algorithm \cite[p. 553]{WLK}.
175,000 iterations of the algorithm were run to produce the
squared-norm-bi-stochastic matrix analyzed (Fig.~\ref{fig:BBSmatrix}). The sum-of-squares difference between the $(3,107)^2$ entries at the 174,999 and 175,000 steps was $4.85 \times 10^{-36}$, indicative of strong convergence. To derive the SK-based bi-stochastic matrix 
(Fig.~\ref{fig:SKmatrix}, \cite[Fig. 6]{SlaterDendrogram}), the particularly simple  procedure presented by Knight \cite[p. 262]{knight} was iterated 115,500 times. No row or column sum of the resultant matrix differed from 1 by more than  $2.94209 \times 10^{-13}$.

Bi-stochastic--or, perhaps more commonly, doubly-stochastic--matrices
are themselves, in general, a subject of major interest in both physical \cite{CSBZ,louck} and mathematical \cite{griffiths,knight,hartfiel}  contexts.

\section{Comparative Analyses}
\begin{figure}
\includegraphics{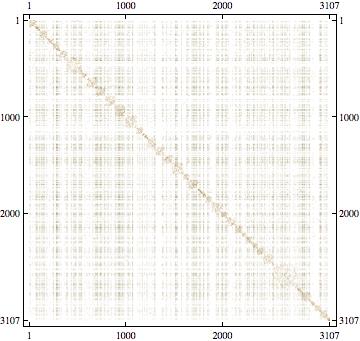}
\caption{\label{fig:BBSmatrix}Bi-stochastic matrix based on minimization of squared-norm Bregman divergence}
\end{figure}

\begin{figure}
\includegraphics{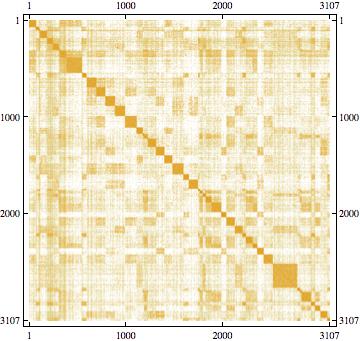}
\caption{\label{fig:SKmatrix}Bi-stochastic matrix--minimizing the Kullback-Leibler form of the Bregman divergence--based on SK-algorithm}
\end{figure}

The correlation coefficient between the two bi-stochastic matrices generated (Figs.~\ref{fig:BBSmatrix} and \ref{fig:SKmatrix} )
is  0.54545. The correlations between the original (raw) intercounty migration matrix (Fig.~\ref{fig:MigrationMatrix}) and Fig.~\ref{fig:BBSmatrix} and Fig.~\ref{fig:SKmatrix} are roughly the same, that is is 0.186802 and 0.176193, respectively.

The leading nine eigenvalues of the 
SK-bi-stochastic matrix are
\begin{equation}
\{1.,0.906253,0.868784,0.84562\, \pm 0.000906366 i,
\end{equation}
\begin{displaymath}
0.831908,0.822294,0.818395\, \pm 0.00243169 i\},
\end{displaymath}
while for the squared-norm-bi-stochastic matrix, these are 
\begin{equation}
\{ i,-i,i,- i,1,1,-1,-0.5 \pm 0.866025 i\}.
\end{equation}
(The eigenvalues of any bi-stochastic matrix can not exceed 1 in absolute value.)

The input (intercounty [destination-origin] migration) matrix--as reported by the U. S. Census Bureau--is "hollow" (zero-diagonal) \cite{dhillon}, so the SK-bi-stochastic matrix must necessarily be also. \footnote{Non-zero diagonal entries might be used to indicate either the number of people moving {\it within} the county, or the [larger] number of people simply still residing [movers {\it and} non-movers] there.} The squared-norm-bi-stochastic matrix, on the other hand, does have a relatively small number (30) of  non-zero (which we take to be those larger than than $10^{-10}$) diagonal 
entries. The greatest is 
0.18361 (Canadian County, Oklahoma), while the second largest is 0.0850351 (Dorchester County, South Carolina) and third largest, 0.0360413 (Onslow County, North Carolina). \footnote{Possible explanatory factors are the presence of prisons in the first two counties, and a large military installation--Camp Lejeune--in the third.} The sum of the 3,107 diagonal entries in this bi-stochastized table is only 0.584398, constituting 
some 0.0188\% of the total of all entries in the table. (One might add a zero-diagonal constraint to the algorithm, but, in this specific case, it seems that it would have negligible effect.)

The original data matrix has 735,531 non-zero entries (being $92.38\%$ sparse), and so then (under broad conditions \cite{knight,hartfiel}) does the SK-bi-stochastic matrix. Contrastingly, the 
squared-norm-bi-stochastic matrix has 57,153 non-zero entries--28,944 of which are greater than  $10^{-10}$.
Most strikingly, the SK-bi-stochastic matrix has only a single entry (corresponding to the flow from the Hawaiian county of Kauai to the Hawaiian county of Kalawao--the second least-populated county in the U. S.) equal to 1, while the squared-norm-matrix has 2,707 such entries, within the high computational accuracy employed. 

\begin{figure}
\includegraphics{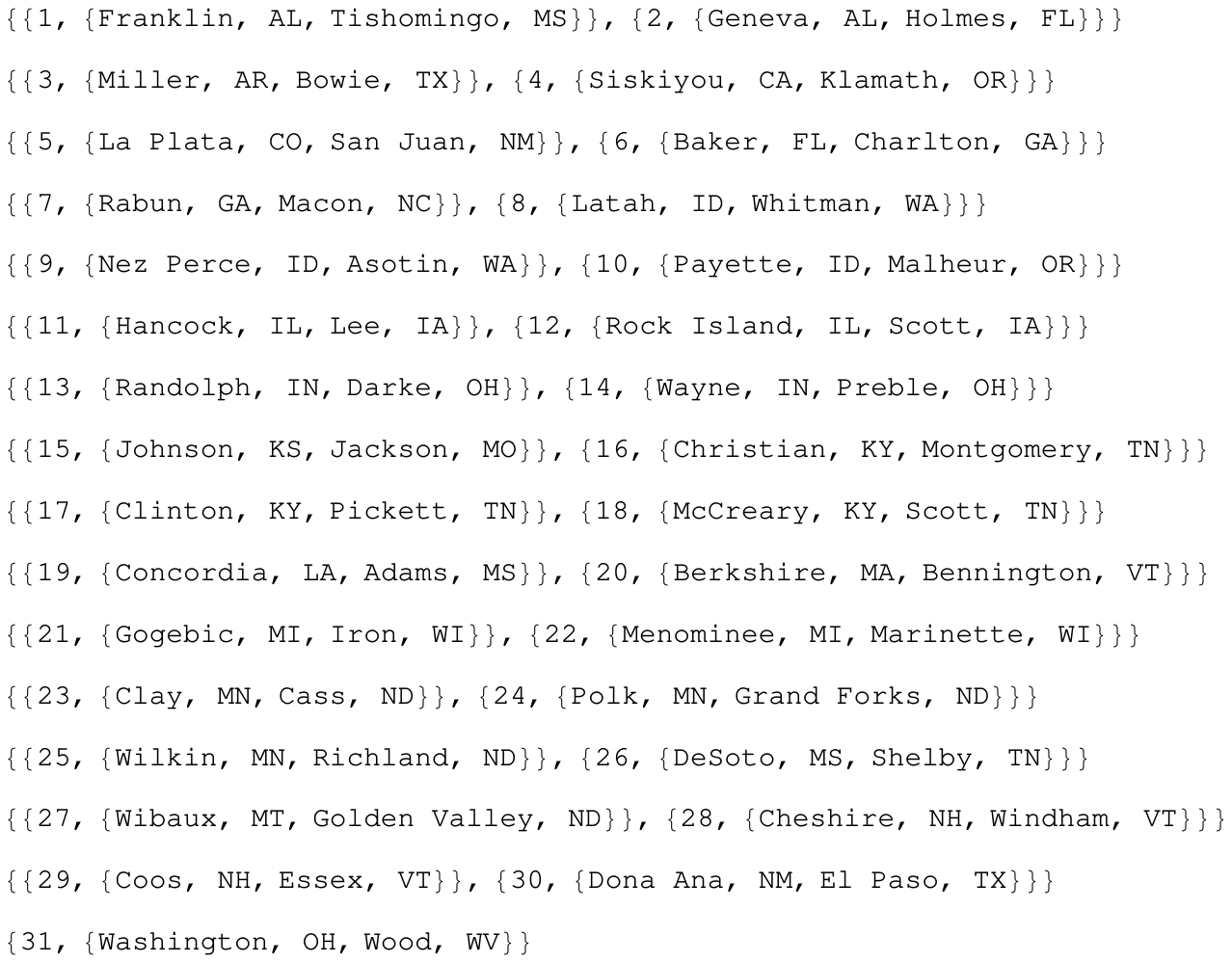}
\caption{\label{fig:Figinterstatepairs}Strong component (interstate) doublets, members lying within different states}
\end{figure}

\begin{figure}
\includegraphics{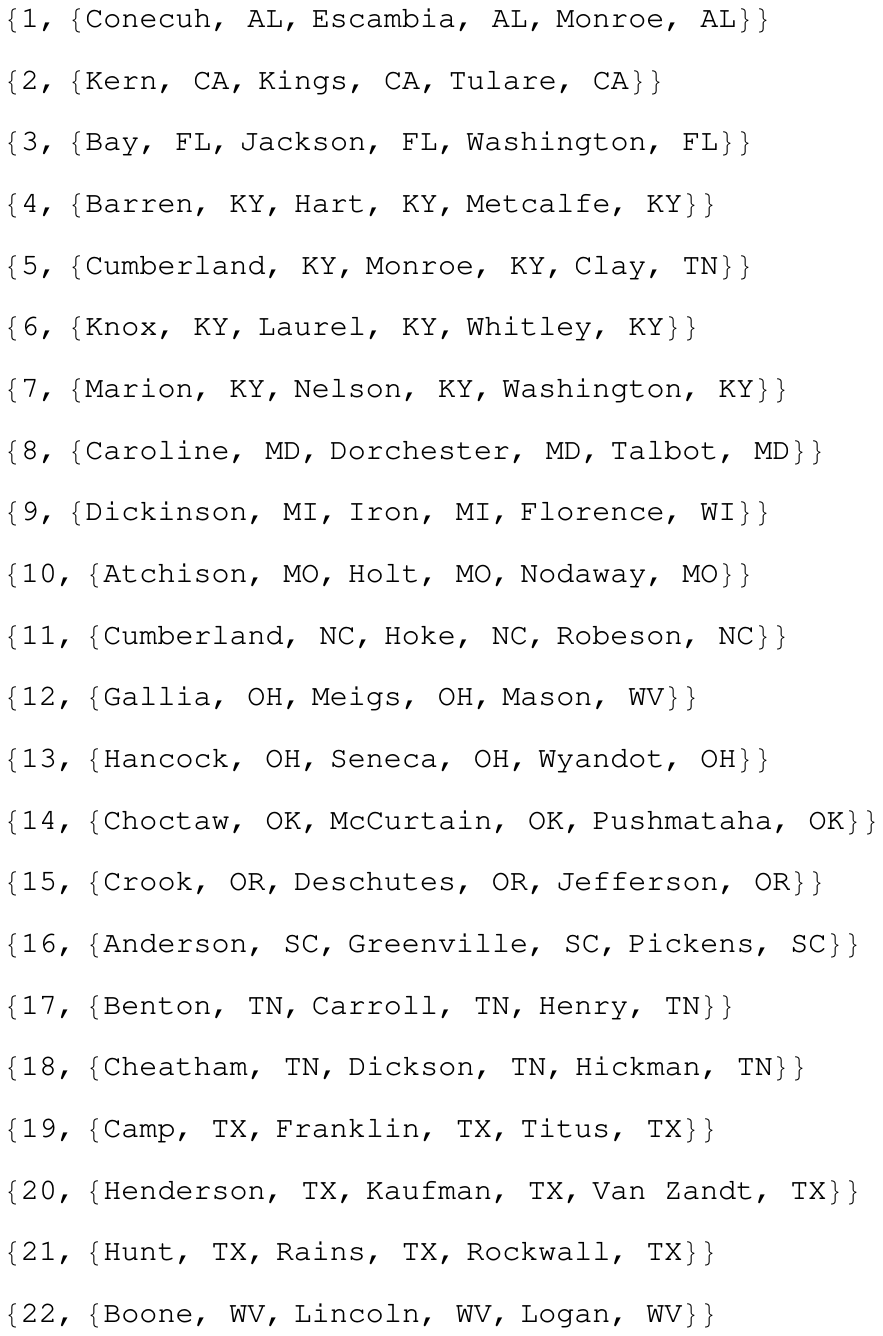}
\caption{\label{fig:FigTriplets}Strong component triplets. The only interstate (Ohio-West Virginia) triplet is the twelfth listed. Mason County, West Virginia and Gallia County, Ohio comprise the  Point Pleasant Micropolitan Statistical Area, as defined by the United States Census Bureau.}
\end{figure}

\begin{figure}
\includegraphics{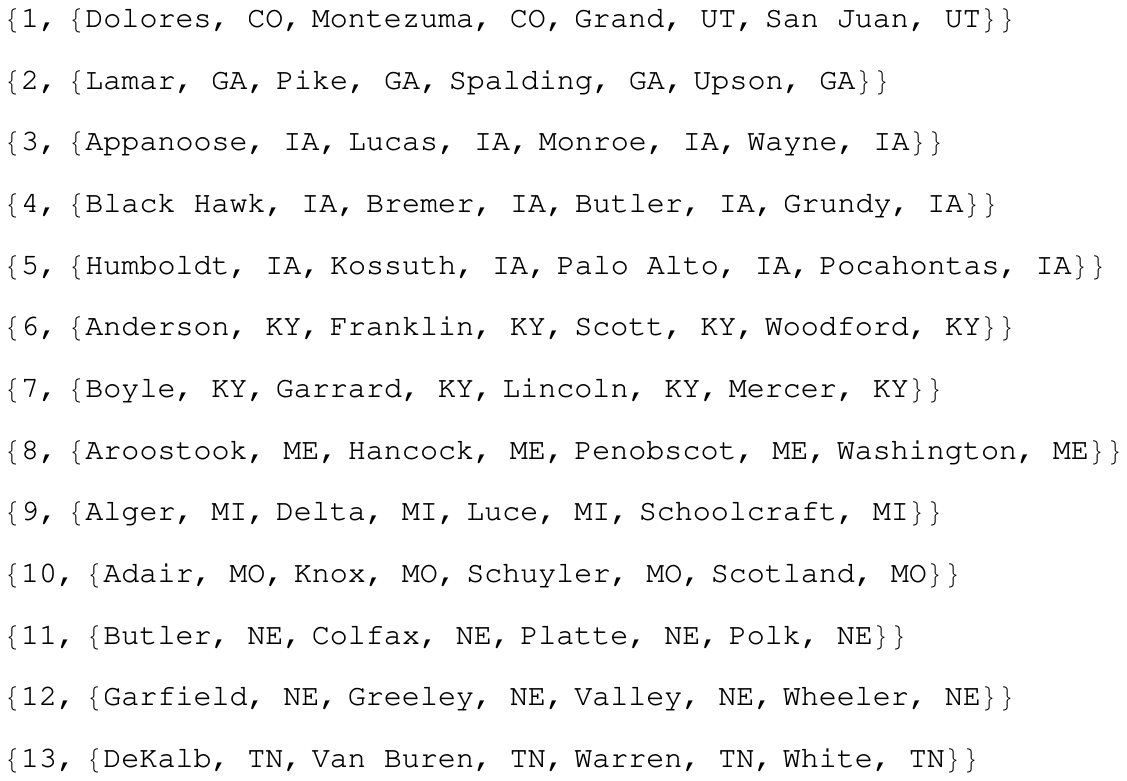}
\caption{\label{fig:FigQuartets}Strong component quartets. The first quartet is composed of two Colorado and two
Utah counties, while the other twelve are all intrastate in nature.}
\end{figure}

\begin{figure}
\includegraphics{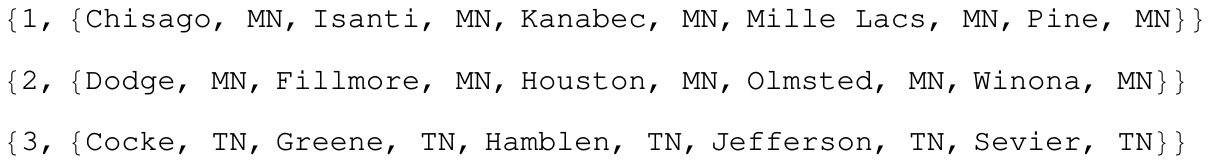}
\caption{\label{fig:FigQuintets}Strong component quintets.
The second quintet listed lies in the Southeast corner 
of Minnesota, while the first quintet is in the East Central part
of the state. The last quintet lies in the eastern part of Tennessee.}
\end{figure}

\begin{figure}
\includegraphics{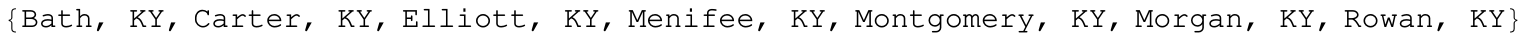}
\caption{\label{fig:FigSeptet}Strong component septet}
\end{figure}

\begin{figure}
\includegraphics{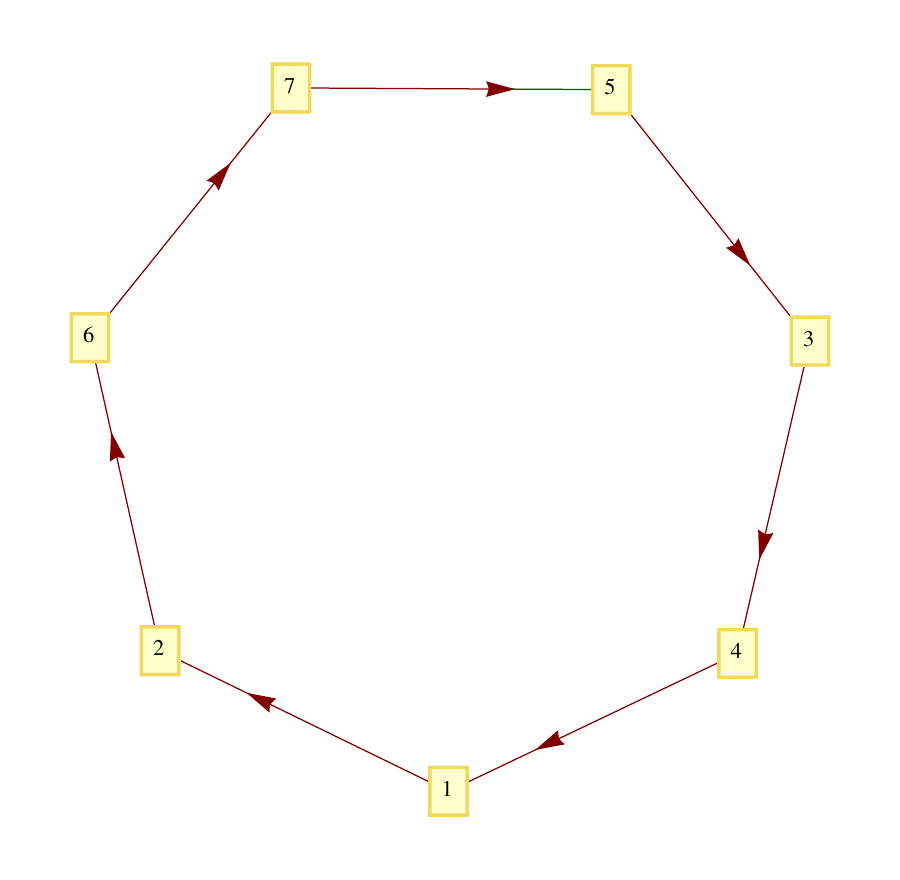}
\caption{\label{fig:KentuckyCounties}Strong component structure
of (eastern) Kentucky septet. Numbering corresponds to the order in which the counties are (alphabetically) listed}
\end{figure}

The corresponding 3,107-vertex, 2,707-link {\it directed} graph possesses 2,352 strong components,  consisting, partly, of  654 pairs--31 of them interstate in nature \footnote{The third interstate pair listed corresponds to the Texarkana, TX-Texarkana, AR Metropolitan Statistical Area (MSA); the eighth to the Lewiston MSA; the tenth to the Ontario Micropolitan Statistical Area; the twelfth to part of the  Davenport-Moline-Rock Island MSA; and the fifteenth to part of the Kansas City metropolitan area. Further, the sixteenth pair comprises part of the 
Clarksville metropolitan area, while  the nineteenth, the Natchez Micropolitan Statistical Area; the twenty-third, the Fargo-Moorhead MSA; the twenty-fourth, the Grand Forks MSA; and the twenty-fifth, the Wahpeton Micropolitan Statistical Area. The twenty-sixth and thirty-first pairs are contained in the Memphis MSA and the ParkersburgÐ-MariettaÐ-Vienna metropolitan area, respectively.} (Fig.~\ref{fig:Figinterstatepairs}) and  623 (Supplemental Material sec.~B), intrastate in character.  There are also 
22 three-member strong components (Fig.~\ref{fig:FigTriplets}, one being interstate), 13 quartets (Fig.~\ref{fig:FigQuartets}, one being interstate), three quintets 
(Fig.~\ref{fig:FigQuintets}), and one septet (Fig.~\ref{fig:FigSeptet}). 
(Fig.~\ref{fig:KentuckyCounties} shows the underlying strong component structure of the Kentucky septet.)
Further, there are 1,659 isolated (ungrouped/non-clustered) counties--959 of which have an entry of  1 in both their row and column (Supplemental Material sec.~C), 600, having a 1 in either their row or column, but not both (Supplemental Material sec.~D) and 100, having no entry of  1 in either their corresponding row and column (Supplemental Material sec.~E).

\begin{figure}
\caption{\label{fig:Page2Dendrogram}SK-based (strong-component hierarchical clustering) dendrogram page containing 5-county Hawaii and 8-county Connecticut clusters}
\includegraphics[page=2,scale=0.9]{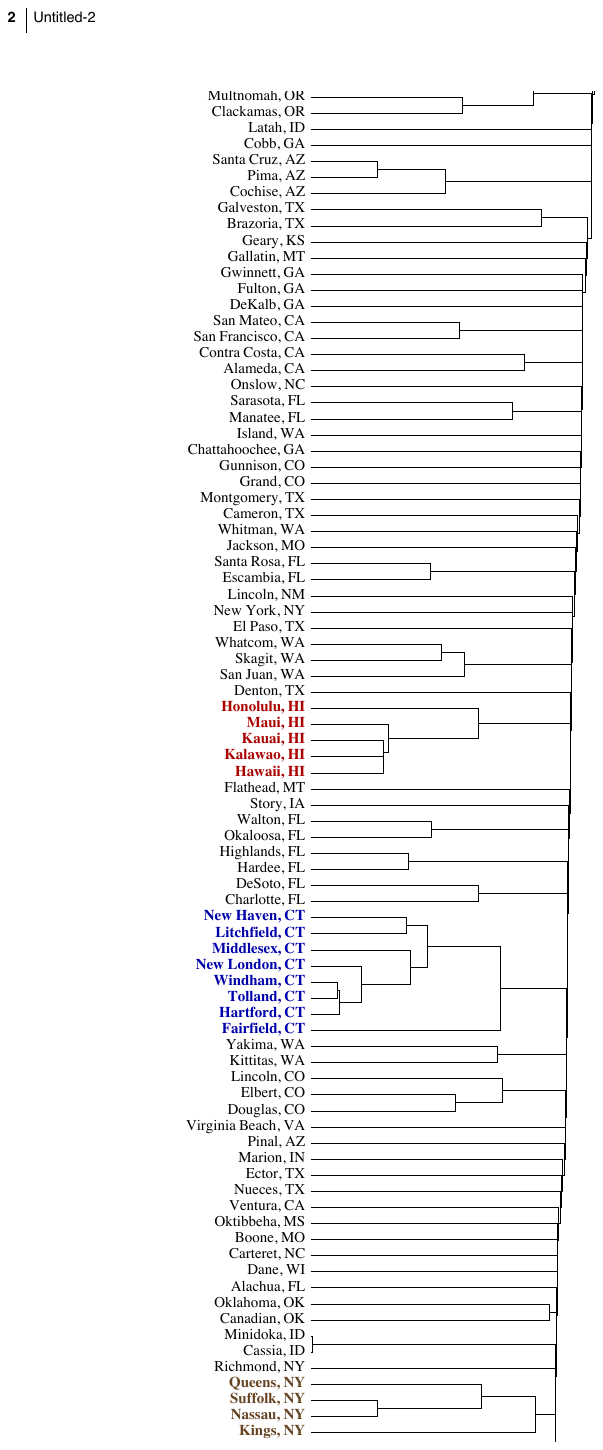}
\end{figure}

\begin{figure}
\caption{\label{fig:Page3Dendrogram}SK-based (strong-component hierarchical clustering) dendrogram page containing 5-county Rhode Island cluster}
\includegraphics[page=3,scale=0.9]{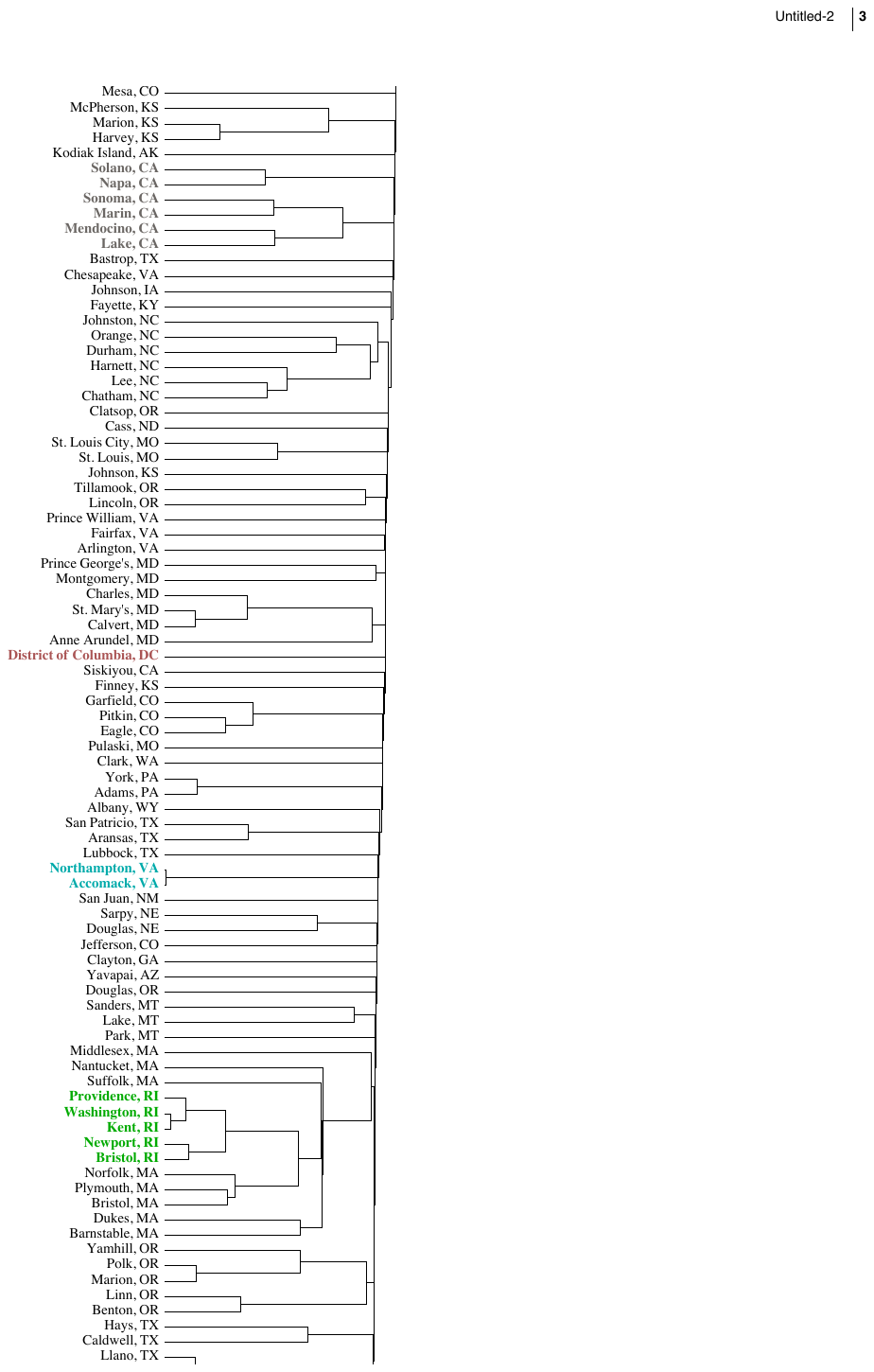}
\end{figure}

Employing the SK-bi-stochastic matrix (Fig.~\ref{fig:SKmatrix}) as input, we had been able to obtain an interesting associated (strong-component hierarchical clustering-based \cite{tarjan2}) dendrogram 
\cite{DendrogramRegionalization}--in line with the extensively-applied two-stage algorithm 
\cite{PBSPNAS,dubes}. However, with the squared-norm-bi-stochastic matrix (Fig.~\ref{fig:BBSmatrix}), such an approach does not appear to be a particularly promising/insightful one--due, it seems, to the extreme skewness of the entries and the fact that the corresponding (0,1) adjacency matrix, based on all 57,153 strictly non-zero entries, is not strongly-connected (having some 694 strong components). 
The five-county states of Hawaii and Rhode Island and the eight-county state of Connecticut--along with other familiar, interesting regional structures \footnote{For example, all the counties of the New England states of Maine, New Hampshire and Vermont, along with two contiguous ones of Massachusetts, formed a 42-county region 
\cite[pp. 9-10]{DendrogramRegionalization}.}--appear (Figs.~\ref{fig:Page2Dendrogram} and \ref{fig:Page3Dendrogram}) as well-defined (hierarchical) clusters in the SK-based (ordinal-scale) dendrogram 
\cite{DendrogramRegionalization}. These three (reconstituted) states are, however, not evident in the  results  based on the bi-stochastic matrix obtained by minimization of the squared-norm form of the Bregman 
divergence. \footnote{In  Supplemental Material sec.~B, there are two pairs 
each of Connecticut, Hawaii and Rhode Island counties--numbered, respectively, (58,59), (100,101) and (456,457). The Rhode Island pair (456)--consisting of Bristol and Newport Counties--is also paired in the dendrogram. The single unpaired remaining [fifth] Hawaii and Rhode Island counties are found in Supplemental Material sec.~E, which also contains one of the four unpaired Connecticut counties.} 

We, further, note that the 3,107-vertex, 2,707-link digraph analyzed above has 1,093 {\it weak} components--the largest containing 29 counties, four from Kentucky and twenty-three from Louisiana and two from Mississippi (Supplemental Material sec.~F). The last two states are contiguous, but neither is adjacent to the first. The second largest weak component of size 27 is intermountain in nature--containing four Idaho, six Washington, seven Nevada, nine Oregon and one California county 
(Supplemental Material sec.~G). The third and fourth largest weak components of size 23 and 21, consist of four Oklahoma and nineteen/seventeen Texas counties (Supplemental Material sec.~H) and eight Arkansas and fifteen Mississippi counties (Supplemental Material sec.~I), respectively. 

The states of Rhode Island, Hawaii and Connecticut do not coincide with any of the 1,093 weak components--as they did not with any of the 2,352 strong components. (Of course, if the digraph under analysis were symmetric, these two numbers would be equal.)

\begin{figure}
\caption{\label{fig:Page1Dendrogram}First page of SK-based (strong-component hierarchical clustering) dendrogram, showing the most (generally Sun-Belt) "cosmopolitan"/hub-like counties and/or groups of counties in descending order from the top}
\includegraphics[page=1,scale=0.85]{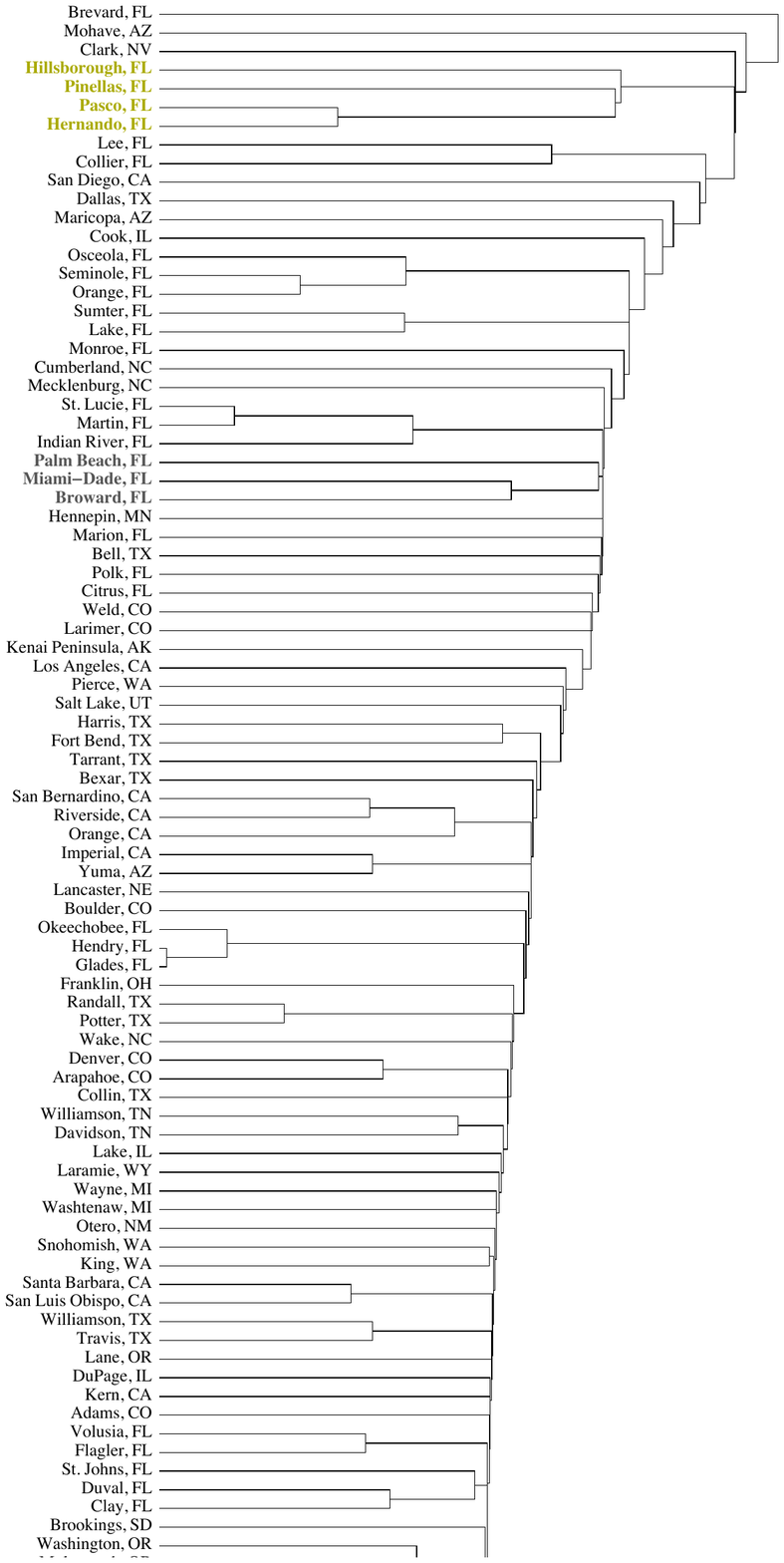}
\end{figure}

In one early application  of the two-stage algorithm \cite{PBSPNAS,dubes} to French interregional migration \cite{France}, the Paris region entered the dendrogram at the far weakest level. Such behavior was deemed in the numerous other national cases subsequently studied to characterize "cosmopolitan" areas (cf. \cite{kleinberg}), those for which the entries in the corresponding SK-based bi-stochastic table were relatively uniformly distributed over the associated row and column. In the U. S. intercounty analysis, the most cosmopolitan/hub-like counties were the Sun-Belt ones of Brevard, FL, Mohave, AZ and Clark, NV (Las Vegas) 
(Fig.~\ref{fig:Page1Dendrogram}). 
\section{Concluding Remarks}
Brockmann and Theis \cite{dirk1} have studied a $3,109 \times 3,109$ U. S. intercounty flow matrix of one-dollar bills. They employed hierarchical clustering  when the network was aggregated to the state level. (No normalization/adjustment  of the raw flows was performed.) They observed that "travel clusters also correspond to geographical clusters" and "geospatial coherence and neighborhood relationships are essentially encoded in the network flow's topology". They interpreted the clusters as "historically grown natural communities because they essentially consist of the west, the northeast, the center, and central east" \cite[p. 33]{dirk1}.
Further, it has recently been emphasized that administrative units--such as the U. S. counties employed above--having developed historically for a diversity of reasons--may not best reflect a natural/functional  contemporary partitioning of a nation into relevant subunits \cite{EEBL,dirk2}.

The squared-norm-bi-stochastization problem is in the nature of a {\it quadratic programming} problem \cite[sec. II]{WLK}. Dorigo and Tobler showed that a certain "push-pull" model of migration, incorporating many of Ravenstein's famous "Laws of Migration", was equivalent to a quadratic transportation problem \cite{DT}. However, interareal distances is an explicit variable in the Dorigo-Tobler treatment, but plays no direct role in the migration-related analyses conducted here (cf. \cite{EEBL}).

Powers of bi-stochastic matrices are also bi-stochastic (smoother in nature)--and, thus, perhaps worthy of further cluster-analytic study (cf. \cite[Fig. 12]{SlaterDendrogram}).

The two forms of Bregman divergence employed above--in our extension of  the investigation of Wang, Li and K{\"o}nig \cite{WLK} to non-symmetric matrices--are the first two listed in Table 2.1 of \cite{dhillon}. However, there are seven additional possible forms of Bregman divergence given--all, presumably, having some possible relevance to the types of questions addressed here. (In \cite[sec. 2.3]{WLK2}, the specific examples of the Mahalanobis and Itakura-Saito distances are commented upon. We are presently investigating the use of the latter form of distance in our U. S. intercounty migration context.)

A chief observation of this study is that a remarkably large percentage, $87.1284\%$ 
(that is, $\frac{2707}{3107}$) of the "mass" of the squared-norm-bi-stochastic table is concentrated on (maximal) unit entries. Further, a quite substantial part, $42.098\%$ (that is,  $\frac{654 \times 2}{3107}$) of the total mass is assigned to maximally 
reciprocally-linked pairs. The SK-based procedure, on the other hand, does not particularly "value" such unit entries (which correspond to  the greatest degree of [one-way] clustering possible). Other forms of Bregman divergences might lead to varying emphases on the production of unit entries. It would appear that the squared-norm-based procedure is, in some sense, "aspiring" to achieve a bi-stochastic table that has the form of a {\it permutation} matrix (decomposable into cycles)--possessing a single unit entry in each row and column.


\section{Supplemental Material}

\includegraphics[page=1]{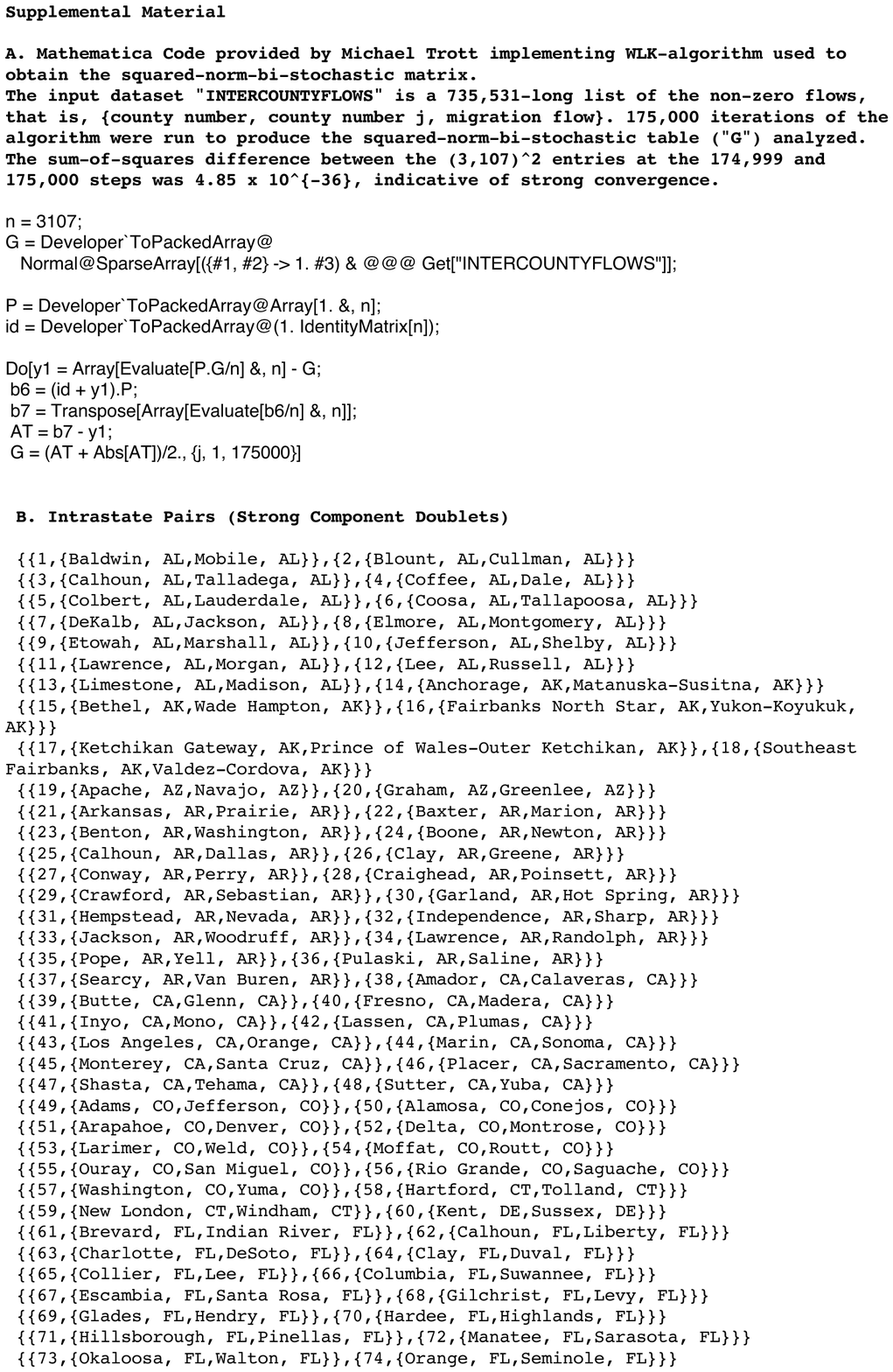}

\includegraphics[page=2]{SupplementalMaterial.pdf}

\includegraphics[page=3]{SupplementalMaterial.pdf}

\includegraphics[page=4]{SupplementalMaterial.pdf}

\includegraphics[page=5]{SupplementalMaterial.pdf}

\includegraphics[page=6]{SupplementalMaterial.pdf}

\includegraphics[page=7]{SupplementalMaterial.pdf}

\includegraphics[page=8]{SupplementalMaterial.pdf}

\includegraphics[page=9]{SupplementalMaterial.pdf}

\includegraphics[page=10]{SupplementalMaterial.pdf}

\includegraphics[page=11]{SupplementalMaterial.pdf}

\includegraphics[page=12]{SupplementalMaterial.pdf}

\includegraphics[page=13]{SupplementalMaterial.pdf}

\begin{acknowledgments}
I would like to express appreciation to the Kavli Institute for Theoretical
Physics (KITP)
for computational support in this research. Michael Trott provided invaluable Mathematica coding assistance (Appendix). Ping Li suggested the topic of this research, and was otherwise encouraging/communicative as well.
\end{acknowledgments}

\bibliography{WangLiKonig4}

\end{document}